\documentclass[prb,preprint]{revtex4-1} 
\usepackage{amsmath}
\usepackage{amsfonts}
\usepackage{physics}
\usepackage{graphicx}
\usepackage{hyperref}

\begin{document}

\title{A Simple Electronic Circuit Demonstrating Hopf Bifurcation for an Advanced Undergraduate Laboratory}

\author{Ishan Deo}
\author{Krishnacharya Khare}
\email{kcharya@iitk.ac.in}
\affiliation{Department of Physics, Indian Institute of Technology Kanpur, Kanpur 208016, India}

\begin{abstract}
A nonlinear electronic circuit comprising of three nodes with a feedback loop is analyzed. The system has two stable states, a uniform state and a sinusoidal oscillating state, and it transitions from one to another by means of a Hopf bifurcation. The stability of this system is analyzed with nonlinear equations derived from a repressilator-like transistor circuit. The apparatus is simple and inexpensive, and the experiment demonstrates aspects of nonlinear dynamical systems in an advanced undergraduate laboratory setting.
\end{abstract}

\maketitle

\section{Introduction}
A simple electronic circuit with a nonlinear electronic component can exhibit chaotic behavior.\cite{Strogatz,Heinrich,Sprott} Mishina et al. used a simple analog electronic circuit and demonstrated successive bifurcations leading to chaotic regimes\cite{Mishina}. Later Hellen et al. used a slightly improved electronic circuit based on a simple nonlinear system, a finite difference equation with a quadratic return map, and showed bifurcation diagrams on an oscilloscope\cite{Hellen}. Goswami et al. used a simple electronic circuit with a voltage-controlled current source which exhibits feature-rich dynamics and several bifurcations\cite{Goswami}. Sack et al. constructed an electronic circuit which possessed remarkable organization in periodic oscillations in parameter space\cite{Sack}. Cabeza et al. developed an electronic circuit with broad cycles\cite{Cabeza}. Even an electronic circuit with a single transistor can display nonlinear behavior like limit cycles, such as in the Hartley and Colpitts oscillators.\cite{Sedra-Smith}

The Hopf bifurcation is a special kind of bifurcation that involves the transition from stable behaviour of a system to periodic and vice versa\cite{Strogatz}. It is prevalent in many physical phenomena, including repressilators in genetic networks, inverted pendulums, fluid flow about a sphere, and electronic circuits with negative differential resistance components\cite{Verdugo, Blackburn, Sharpe, Drazin, Wallis}. Particularly, it is from the repressilator-based configuration that we get our motivation for the circuit. A repressilator is a genetic regulatory network comprised of at least one feedback loop of at least three genes. Taking inspiration from this, Rim et al. constructed an electronic circuit with a negative feedback loop consisting of three nodes, where each node models a gene\cite{Rim}. The primary component of each node is a transistor, which serves as the means for negative feedback between nodes. We present a detailed analysis of the dynamics of the Hopf bifurcation in this $3$--node electronic circuit. We obtain explicit formulae for the dependence of the bifurcation resistance on the applied voltage and for the variation of the fixed point voltage with resistance and applied voltage. The transistors used here are inexpensive, and the other components are easily found in any undergraduate physics laboratory. The theory of the Hopf bifurcation is also relatively simple, requiring only knowledge of multivariable calculus and calculus-based classical mechanics. Coupled with the simplicity of the circuit design and the circuit analysis, this experiment is appropriate to introduce advanced undergraduate students to the fascinating world of nonlinear dynamics.

\section{Structure of the Circuit}
\begin{figure}[h!]
    \centering
    \includegraphics[width=8.5cm]{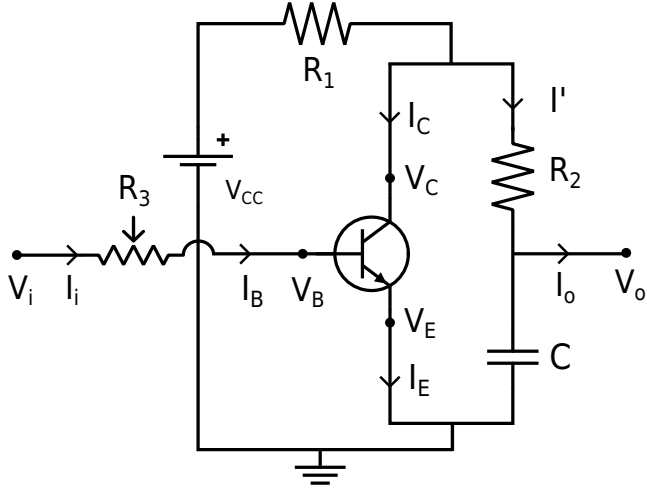}
    \caption{Circuit diagram of one node showing the components and parameters. $R_3$ is the variable input resistor which is also used as the feedback resistor when different nodes are connected together.}
    \label{Node}
\end{figure}
\begin{figure}[h!]
    \centering
    \includegraphics[width=16cm]{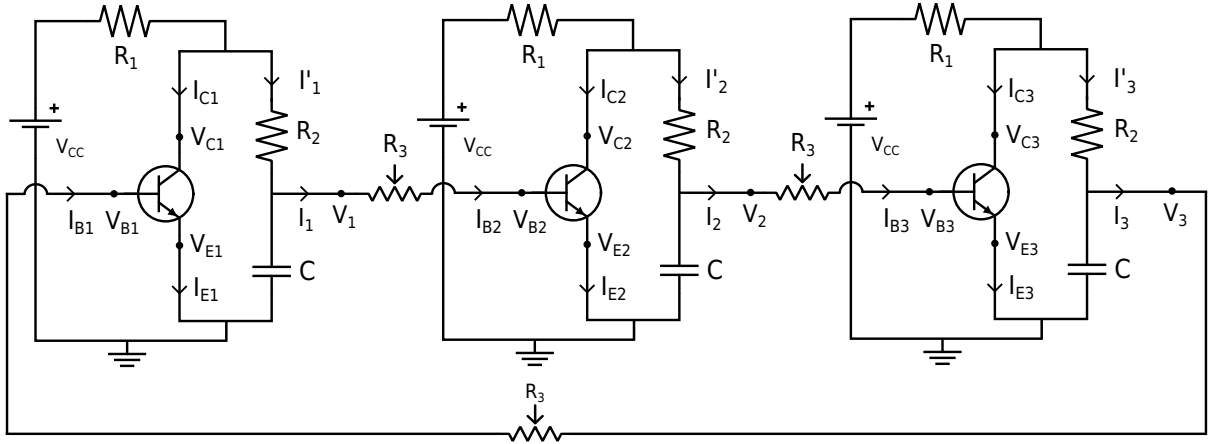}
    \caption{Circuit diagram of the complete circuit having $3$ nodes and feedback loop.}
    \label{Circuit}
\end{figure}

The electronic circuit we use here has been taken from the paper by Rim et al\cite{Rim}. It consists of three identical nodes, which are connected to each other in a cycle to form a negative feedback loop. The structure of each node is shown in Fig. \ref{Node}, where the transistor, resistors $R_1$ and $R_2$, and the capacitor all have fixed values. The resistor $R_3$ is variable, and is the independent bifurcation parameter of the system. The transistor used in a node is an NPN Bipolar Junction Transistor (BJT) with $V_B, V_C$ and $V_E$ as base, collector and emitter voltages respectively and $I_B, I_C$ and $I_E$ as corresponding currents. $V_i, I_i, V_o$ and $I_o$ represent input and output voltages and currents respectively and $V_{CC}$ is the supply voltage. In the analysis, all the time-dependent parameters are written in corresponding lower case letters. The entire circuit comprising all the three nodes is shown in Fig. \ref{Circuit}. So, the cycle is formed by: 
\begin{itemize}
    \item The $v_o$ of the $1^\mathrm{st}$ node serving as the $v_i$ of the $2^\mathrm{nd}$ node
    \item The $v_o$ of the $2^\mathrm{nd}$ node serving as the $v_i$ of the $3^\mathrm{rd}$ node
    \item The $v_o$ of the $3^\mathrm{rd}$ node serving as the $v_i$ of the $1^\mathrm{st}$ node
\end{itemize}
In the whole loop, each of the three resistances labelled $R_3$ have the same values at all times, and are varied simultaneously.

\section{Circuit Analysis} \label{Analysis}
\subsection{Modelling of an NPN BJT in Active Mode}
\label{Ebers-Moll}
An NPN BJT has four modes of operation: active, cutoff, reverse active, and saturation, which depend on whether $v_{BE} = v_B-v_E$ and $v_{CB} = v_C-v_B$ are positive or negative, where $v_{BE}$ and $v_{CB}$ are the transient voltages at the base-emitter and collector-base junctions respectively. In our experimental setup, BJT is in the active mode (that is, both $v_{BE}$ and $v_{CB}$ are positive). So, we focus on this mode only here. For an NPN BJT in the active mode, the Ebers-Moll model gives, under the assumption of physically realistic voltages, the equations
\begin{align}
    i_C &= I_S\exp\frac{v_{BE}}{V_T}, \\
    i_C &= \alpha i_E, \\
    i_C &= \beta i_B.
\end{align}
Here, $I_S$ is the saturation current, $V_T$ the thermal voltage, and $\alpha$ and $\beta$ are two proportionality constants. Now, we also have $i_E = i_C + i_B$, with all three positive (as $v_{BE}$ and $v_{CB}$ are positive). Thus, we get the relations $\alpha=\beta/(1+\beta)$ or $\beta=\alpha/(1-\alpha)$ between $\alpha$ and $\beta$. For typical NPN BJTs, $\beta$ ranges from $50$ to $200$, and so $\alpha$ ranges from $0.98$ to $0.99$. Also, $I_S$ is of the order of $10^{-12} A$ for typical NPN BJTs, and $V_T = 0.025875 V$ at room temperature ($298 K$). For further details on NPN BJTs, readers are encouraged to consult Ref. [9].\cite{Sedra-Smith} 

\subsection{Lambert $W$ Function}
The Lambert $W$ function $W(x)$ is defined as the real solution of the equation $ye^y = x$. We summarize here some properties of this function for positive $x$, which will be used throughout. For detailed proofs, one may look at Ref. [17].\cite{Stewart}
\begin{itemize}
    \item\textbf{Single-Valued:} The function $W(x)$ is single valued, that is, for any $x$, there is only one real $y$ such that $ye^y = x$.
    \item\textbf{Positive:} For $x>0$, $W(x)>0$.
    \item\textbf{Monotonic:} $W(x)$ is a monotonically increasing function. $\\$ That is $x_1>x_2 \implies W(x_1) > W(x_2)$.
    \item\textbf{Derivative:} 
    \begin{equation}
        \dv{W(x)}{x} = \frac{1}{x}\frac{W(x)}{1+W(x)}.
    \end{equation}
\end{itemize}

\subsection{Modelling the Node}
For this circuit, as the emitter is grounded in the node, $v_E = 0$, and so $v_B = v_{BE}$. Applying Kirchoff's Laws and the results in Section \ref{Ebers-Moll}, we get
\begin{align}
    \dv{v_o}{t} &= \frac{V_{CC}-v_o-i_CR_1-i_o(R_1+R_2)}{C(R_1+R_2)} \label{eq:timevar} \\
    \log\frac{i_C}{I_S} &= \frac{v_i}{V_T}-\frac{R_3}{\beta V_T}i_C.
\end{align}
Let us take $a = I_SR_3/\beta V_T$ for simplicity. Then, we can solve for $i_C$ with the help of the Lambert W function $W(x):$
\begin{equation}
    i_C = \frac{\beta V_T}{R_3}W\left(a\exp\frac{v_i}{V_T}\right).
\end{equation}
Note that $a$ depends only on $R_3$ (the bifurcation parameter). On substituting this in Eq. \ref{eq:timevar}, we get the variation of $v_o$ as
\begin{equation}
    \dv{v_o}{t} =\frac{V_{CC}-v_o-\beta V_T\frac{R_1}{R_3} W\left(a\exp\frac{v_i}{V_T}\right)-i_o(R_1+R_2)}{C(R_1+R_2)}.
\end{equation}

\subsection{Modelling the Circuit}
For this circuit, the input of a node is the output of the previous node. So, the values of $V$ and $I$ of a node depend on both the preceding and succeeding nodes. Now, formally, let us define node $4$ to be identical to node $1$, and node $0$ as identical to node $3$. Then, we can say that node $n$ depends on the nodes $n-1$ and $n+1$. Thus, we get for current
\begin{align}
    i_{C_n} &= \frac{\beta V_T}{R_3}W\left(a\exp\frac{v_{n-1}}{V_T}\right), \\
    i_n &= i_{B_{n+1}} \\
    &= \frac{V_T}{R_3}W\left(a\exp\frac{v_n}{V_T}\right),
\end{align}
and hence for voltage
\begin{equation}
    \dv{v_n}{t} = \frac{V_{CC} - v_n}{C(R_1+R_2)} - \frac{V_T}{CR_3} \left[ W\left(a\exp\frac{v_n}{V_T}\right) + \beta \frac{R_1}{R_1+R_2} W\left(a\exp\frac{v_{n-1}}{V_T}\right) \right].
\end{equation}

\section{Mathematical Analysis of the Model}
Due to the nonlinearity, we are unable to find a closed form solution for this system. Instead, we analyze the behaviour of the stable states of the system.

\subsection{Fixed Point}
The system has a fixed point when $\text{d}v_n/\text{d}t = 0$ for $n=1,2,3$. That is
\begin{align}
    V_{CC} &= v_1 + V_T\left[\frac{R_1+R_2}{R_3} W\left(a\exp\frac{v_1}{V_T}\right) + \beta \frac{R_1}{R_3} W\left(a\exp\frac{v_3}{V_T}\right) \right] \label{fp1}; \\
    V_{CC} &= v_2 + V_T\left[\frac{R_1+R_2}{R_3} W\left(a\exp\frac{v_2}{V_T}\right) + \beta \frac{R_1}{R_3} W\left(a\exp\frac{v_1}{V_T}\right) \right] \label{fp2}; \\
    V_{CC} &= v_3 + V_T\left[\frac{R_1+R_2}{R_3} W\left(a\exp\frac{v_3}{V_T}\right) + \beta \frac{R_1}{R_3} W\left(a\exp\frac{v_2}{V_T}\right) \right] \label{fp3}.
\end{align}

\subsubsection{Equality of $v_n$ at Fixed Point} 
Without loss of generality, we can assume that $v_1\le v_2\le v_3$ at the fixed point. Subtracting Eq. \ref{fp2} from Eq. \ref{fp3}, we get
\begin{align}
    &(v_3-v_2) + V_T \frac{R_1+R_2}{R_3} \left[W\left(a\exp\frac{v_3}{V_T}\right) - W\left(a\exp\frac{v_3}{V_T}\right) \right] \nonumber \\
    &= V_T\beta \frac{R_1}{R_3} \left[ W\left(a\exp\frac{v_1}{V_T}\right) - W\left(a\exp\frac{v_2}{V_T}\right)\right].
\end{align}
The function $W(x)$ is monotonic for positive arguments. So, as $v_3\ge v_2$ by assumption, the left hand side of the equation is $\ge 0$. But, as $v_1\le v_2$, the right hand side of the equation is $\le 0$. Thus, the two sides are equal only when $v_1 = v_2 = v_3 \equiv v_P$. Hence, any fixed point of the system must be of the form $(v_P,v_P,v_P)$.

\subsubsection{Existence of Unique Fixed Point}
At a fixed point $(v_P,v_P,v_P)$, the equation of the system is 
\begin{equation}
    V_{CC} = v_P + V_T \frac{R_1(1+\beta)+R_2}{R_3} W\left(a\exp\frac{v_P}{V_T}\right).
\end{equation}
Let $R = R_1(1+\beta)+R_2 and R' = R+R_3$ for simplicity. Then, we can solve for $v_P$ to get
\begin{equation}
    v_P = V_{CC} - V_T\frac{R}{R'} W\left(\frac{I_SR'}{\beta V_T}\exp\frac{V_{CC}}{V_T}\right) \label{eq:V_PvsV_CC}.
\end{equation}
Thus, there exists a fixed point, and as $W(x)$ is single-valued, it is unique. Let us also derive a useful equation for future use. Defining $y = (I_SR'/\beta V_T)\exp (V_{CC}/V_T)$ for ease of notation, we have
\begin{equation}
    W\left(a\exp\frac{v_P}{V_T}\right) = \frac{R_3}{R'} W(y).
\end{equation}

\subsection{Jacobian and its Eigenvalues}
In order to determine whether the system has a Hopf bifurcation, we need to determine the eigenvalues of the Jacobian. The Jacobian of the system has $2$ terms:
\begin{align}
    A &= \left(\pdv{}{v_n}\dv{v_n}{t}\right)\Big\vert_{P} \\
    &= -\frac{1}{C}\left(\frac{1}{R_1+R_2} + \frac{W(y)}{R' + R_3W(y)} \right), \\
    B &= \left(\pdv{}{v_{n-1}}\dv{v_n}{t}\right)\Big\vert_{P} \\
    &= -\frac{\beta}{C}\frac{R_1}{R_1+R_2} \frac{W(y)}{R' + R_3W(y)}.
\end{align}
In terms of these parameters, the Jacobian is 
\begin{equation}
    J = \begin{bmatrix}
    A & 0 & B \\
    B & A & 0 \\
    0 & B & A \\
    \end{bmatrix}.
\end{equation}
This matrix has the eigenvalues
\begin{align}
    \lambda_1 &= A - \frac{B}{2} + \frac{\sqrt 3}{2}Bi, \\
    \lambda_2 &= A - \frac{B}{2} - \frac{\sqrt 3}{2}Bi,\\
    \lambda_3 &= A + B.
\end{align}
As $A$ and $B$ are real, we have a pair of complex eigenvalues.

\subsection{Existence of Hopf Bifurcation}
The mathematics behind Hopf bifurcations is well described in textbooks\cite{Strogatz, Guckenheimer}. Interested readers can consult them to learn about the existing result of the Hopf bifurcation and its proof. In terms of the system we are analyzing, the existing result translates to the following. The circuit displays a Hopf bifurcation if it has a fixed point of the node voltage $(v_P, v_P, v_P)$, and at this fixed point, there exists a critical value $R_c$ of the bifurcation parameter $R_3$ at which the eigenvalues of the Jacobian satisfy the following properties:
\begin{align}
    \text{Re}(\lambda_1) &= \text{Re}(\lambda_2) = 0 \text{ and Re}(\lambda_3) \neq 0 \\
    \dv{}{R_3}\left(\text{Re}(\lambda_1)\right) &= \dv{}{R_3}\left(\text{Re}(\lambda_2)\right) \neq 0 
\end{align}
As we saw in the first subsection of this section, the system does indeed have a fixed point. Also, as $A, B$ are negative, $\lambda_3 = A + B \neq 0$, which satisfies part of the first condition. For the rest, it is evident that the existence of Hopf bifurcation depends on
\begin{equation}
    A - \frac{B}{2} = -\frac{1}{C(R_1+R_2)} \left[1+ \left(R_1+R_2-\frac{\beta R_1}{2}\right) \frac{W(y)}{R' + R_3W(y)} \right].
\end{equation}
Let us check the second condition first. Computing the derivative, we get
\begin{equation}
    \dv{}{R_3}\left(A-\frac{B}{2}\right) = \frac{R_1+R_2-\frac{\beta R_1}{2}}{C(R_1+R_2)} \frac{W(y)^2}{(R'+R_3W(y))^2} \frac{2+W(y)}{1+W(y)}.
\end{equation}
As $W(x)>0$ for $x>0$, this derivative is non-zero for all values of $R_3$ - that is, the second condition holds for any $R_3$. For the first condition, let us define $R_0 = \beta R_1/2-R_1-R_2-R_3$. Then, $A=B/2$ implies
\begin{equation}
    W(y) = \frac{R'}{R_0} = \frac{3\beta R_1}{2R_0}-1.
\end{equation}
After some algebraic manipulations, this becomes
\begin{equation}
    R_3 = \frac{\beta R_1}{2}\left(1-\frac{3}{W\left(\frac{3I_SR_1}{2V_T} \exp\frac{V_{CC}+V_T}{V_T} \right)}\right) - R_1 - R_2.
\end{equation}
Thus, we get a critical value of $R_3$, and hence, the system displays a Hopf bifurcation. For future use, the above equation can be inverted to get $V_{CC}$:
\begin{equation}
    \frac{V_{CC}}{V_T} = \frac{3}{1 -  \frac{2}{\beta} \frac{R_3+R_1+R_2}{R_1} } + \log \frac{\frac{2V_T}{I_SR_1}}{1 -  \frac{2}{\beta} \frac{R_3+R_1+R_2}{R_1} } - 1. \label{eq:Bifurcation}
\end{equation}
From this, as $W(x)>0$ for $x>0$, we also get the inequality
\begin{equation} \label{eq:Ineq}
    \beta > 2 \frac{R_3+R_1+R_2}{R_1}.
\end{equation}
Thus, this inequality holds if and only if the system displays a Hopf bifurcation at that value of $R_3$.

\section{Experimental Setup}
We constructed the circuit as shown in Fig. \ref{Circuit} with the following circuit components:
\begin{itemize}
    \item{$R_1$:} $1$ k$\Omega $ potentiometer
    \item{$R_2$:} $1$ k$\Omega $ potentiometer
    \item{$R_3$:} $100$ k$\Omega $ potentiometer
    \item{$C$:} $220$ nF capacitor
    \item Transistor (NPN) Model: SL100
\end{itemize}
\begin{figure}[h!]
    \centering
    \includegraphics[width=8.5 cm]{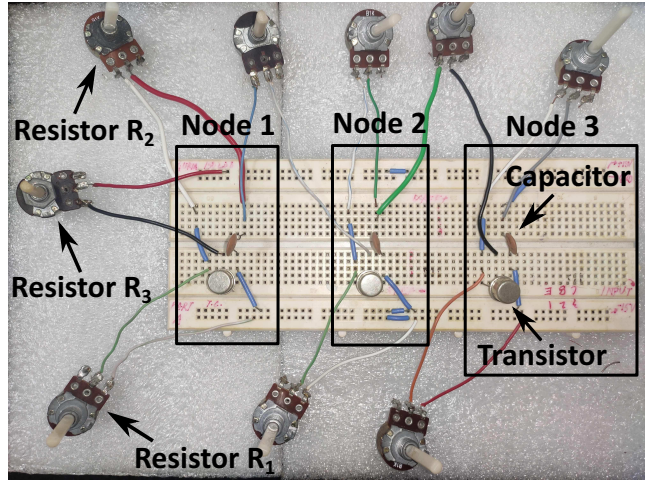}
    \caption{Photograph of the actual circuit prepared on a breadboard showing all three nodes with electronic components.}
    \label{Real_Life}
\end{figure}

Note that even though the resistance values $R_1$ and $R_2$ are held constant during the experiment, we have used potentiometers for them. This is to ensure that the values of $R_1$ and $R_2$ are exactly $1$ k$\Omega$, and thus reduce the error in the voltages measured. We used an ordinary lab digital storage oscilloscope (DSO) to measure the voltages $v_i$ of the capacitor as a function of the different values of resistance $R_3$ and applied voltage $V_{CC}$. As it was easier to change $V_{CC}$ as compared to $R_3$, we measured $v_i$ on varying $V_{CC}$ at a fixed $R_3$, for multiple values of $R_3$.

\section{Results and Discussions}
\subsection{Preliminary Results}
\subsubsection{Verification of Active Mode of Transistors}
In each measurement, for each transistor, the voltage $v_{BE}$ was around $0.65-0.75$ V, while the voltage $v_{CB}$ was positive. So, the transistor is indeed operating in the active mode.

\subsubsection{$\beta$ Value of Transistors}
The $\beta$ value of the transistor of each node is given in Table \ref{table:Beta}.
\begin{table}[h!]
    \centering
    \setlength{\tabcolsep}{4pt}
    \begin{tabular}{|c|c|c|c|c|c|}
    \hline
    Node 1 & Node 2 & Node 3 & Mean & Standard Deviation & \% Error \\
    \hline
    129.6 & 131.7 & 130.2 & 130.5 & 1.08 & 0.83 \\
    \hline
    \end{tabular}
    \caption{$\beta$ Values of Transistors of the Circuit}
    \label{table:Beta}
\end{table}
These values are not equal, possibly due to manufacturing defects. Thus, we use the mean value of $\beta = 130.5$.

\subsection{Existence of Hopf Bifurcation and Variation with Applied Voltage}
\begin{figure}[h!]
    \centering
    \includegraphics[width=8.5cm]{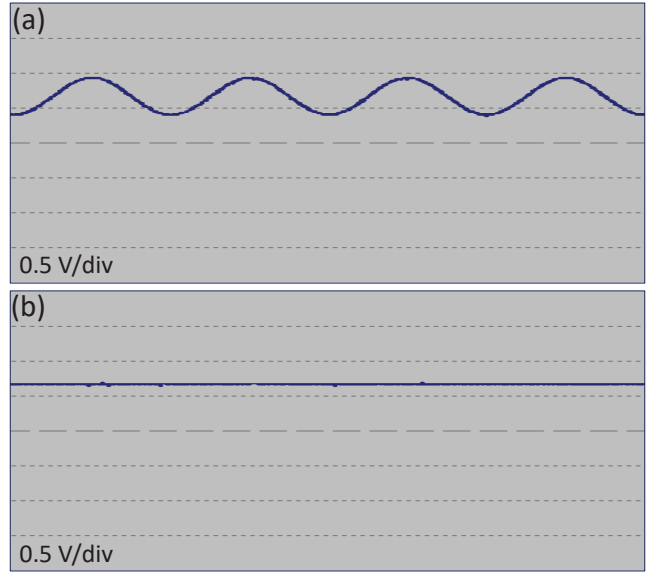}
    \caption{Hopf bifurcation at node $2$ for a bifurcation resistance $R_3 = 61.0$ k$\Omega$ by varying $V_{CC}$. Waveforms in a digital oscilloscope at (a) voltage before the Hopf bifurcation, and (b) afterwards.}
    \label{fig:Waveform}
\end{figure}

\begin{figure}[t!]
    \centering
    \includegraphics[width=8.5cm]{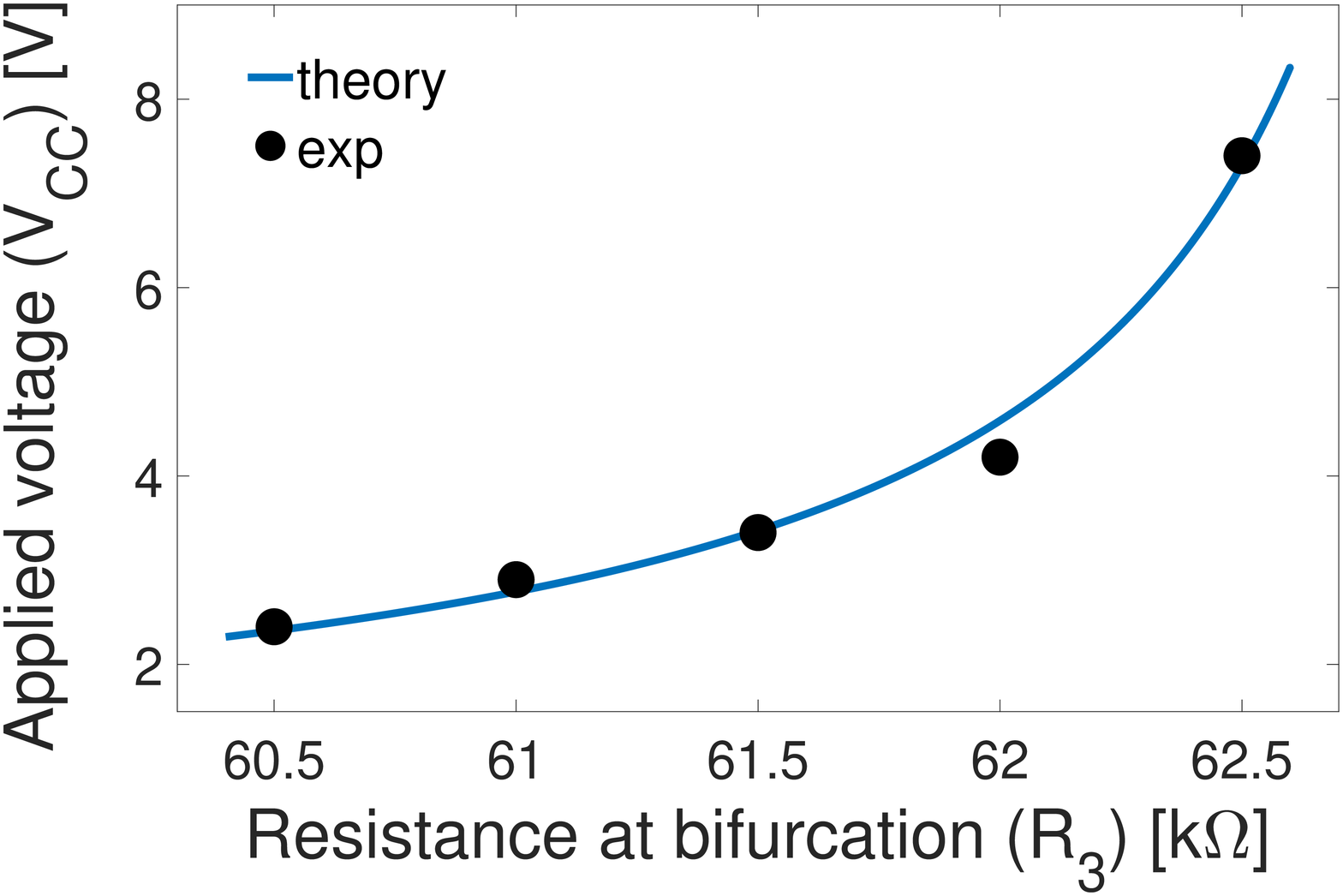}
    \caption{Applied Voltage ($V_{CC}$) versus Resistance at Bifurcation ($R_3$). $R^2$ value of this plot is $0.9887$.}
    \label{fig:Bifurcation}
\end{figure}

On fixing a value of $R_3$, we varied $V_{CC}$ applied to the circuit, in order to find its value at which we have a Hopf bifurcation. Figure \ref{fig:Waveform} shows the Hopf bifurcation on a digital oscilloscope at node 2 of the circuit by varying $V_{CC}$ with $R_3 = 61.0$ k$\Omega$.
The experimental values of $R_3$ and $V_{CC}$ are listed in the Table \ref{table:Bifurcation}. 
\begin{table}[h]
    \centering
    \setlength{\tabcolsep}{4pt}
    \begin{tabular}{|c|c|}
    \hline
    Resistance at Bifurcation ($R_3$) [k$\Omega$] & Applied Voltage ($V_{CC}$) [V]\\
    \hline
    64.0 & No Bifurcation Occurs \\
    \hline
    62.5 & 7.4 \\
    \hline
    62.0 & 4.2 \\
    \hline
    61.5 & 3.4 \\
    \hline
    61.0 & 2.9 \\
    \hline
    60.5 & 2.4 \\
    \hline
    \end{tabular}
    \caption{Bifurcation resistance versus applied Voltage Variation}
    \label{table:Bifurcation}
\end{table}
From the data, we can see that there is no Hopf bifurcation at $R_3 = 64$ k$\Omega$, as we have 
\begin{equation}
    2 \frac{R_3+R_1+R_2}{R_1} = 2\frac{64\times 10^3+10^3+10^3}{10^3} = 132 > \beta = 130.5,
\end{equation}
which satisfies the condition of inequality (Eq. \ref{eq:Ineq}). For smaller values of $R_3$, Hopf bifurcation occurs. In Fig. \ref{fig:Bifurcation}, we have plotted these values and compared them to the theoretical curve. In Fig. \ref{fig:Bifurcation}, we see that the experimental data closely matches with the theoretical curve constructed using Eq. \ref{eq:Bifurcation}. This supports the theoretical result we had derived previously.

\subsection{Type of Hopf Bifurcation}
There are two types of Hopf bifurcations: supercritical and subcritical. The most important difference between them, for us, is reversibility. That is, suppose we vary the bifurcation resistance $R_3$ such that the bifurcation occurs and then go back to the original value of $R_3$. Then, the system displaying supercritical Hopf bifurcation will revert back to the original voltage state, while it will not for a subcritical Hopf bifurcation. On performing this test to the circuit, we observed that for each value of applied voltage, the fixed-point node voltage returned to the original value. Thus, the circuit displays a supercritical Hopf bifurcation.

\subsection{Variation of Fixed Point Voltage with Applied Voltage and Resistance}
\begin{figure}[h!]
    \centering
    \includegraphics[width=8.5 cm]{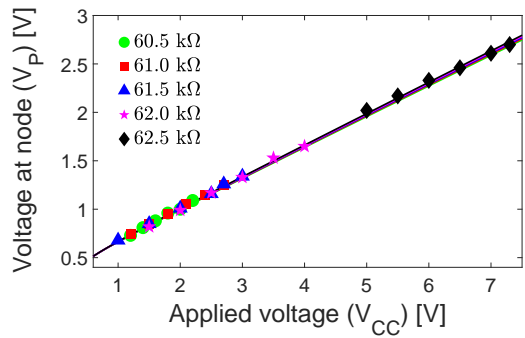}
    \caption{Plots of Fixed Point Voltage ($V_P$) with Applied Voltage ($V_{CC}$) for different Bifurcation Resistances ($R_3$). The solid curve represents the theoretical result given by the Eq. \ref{eq:V_PvsV_CC}, while the data points represent the experimentally obtained values. The $R^2$ values of all 5 graphs range from $0.9753$ to $0.9935$.}
    \label{fig:V_PvsV_CC}
\end{figure}

In the fixed point regime of the ciruit, we varied the resistance $R_3$ and applied voltage $V_{CC}$ and observed the values of $v_P$ obtained at each node. From the readings, we can see that the voltages in each of the three nodes are not identical. This can be attributed to the three nodes being dissimilar, which occurs due to the transistors having different $\beta$ values, for reasons mentioned before. We also plotted the variation of $v_P$ with $V_{CC}$ at a fixed resistance $R_3$, and compared it with the theoretical curve (obtained from Eq. \ref{eq:V_PvsV_CC}) in Fig. \ref{fig:V_PvsV_CC}. We can see that there is an excellent match between experimental data and theoretical results, thus supporting the results we have obtained.

\section{Conclusion}
In this paper, we present a simple electronic circuit consisting of three nodes and a feedback loop, which is similar to a genetic repressilator. The circuit undergoes a Hopf bifurcation based on various system parameters. DC analysis of the circuit tells us how the voltage (amplitude) varies with the resistance (bifurcation parameter) in the limit cycle close to the bifurcation point. Since all components of the apparatus involved are readily available, and prerequisite knowledge is within the level of a $3^{\text{rd}}$ year college education in physics, this experiment serves as a good introduction for students to experimental nonlinear dynamics.

\section{Acknowledgement}
This study was supported by the Modern Physics Laboratory of the Department of Physics, Indian Institute of Technology Kanpur and Science and Engineering Research Board (SERB) New Delhi (Project no. CRG/2019/000915). The authors thank Upendra Kumar Parashar for valuable help with the experiments.

\end{document}